\documentclass[prb,twocolumn,showpacs,amsmath,amssymb]{revtex4}
\usepackage{times}
\usepackage{mathptmx}
\usepackage{dcolumn}
\usepackage{epsf}
\usepackage{graphicx}
\epsfverbosetrue

\def\rmd     {\hbox{\scriptsize d}}
\def\rmC     {\hbox{\scriptsize C}}
\def\rmT     {\hbox{\scriptsize T}}
\def\rmRx    {\hbox{\scriptsize :Rx}}
\def\rmId    {\hbox{\scriptsize :Id}}
\def\Or      {\hbox{O}\!}

\def\nl	  	 { \hfill\break }

\def\VI      {$\hbox{V}_{\hbox{\scriptsize In}}$}
\def\VIs     {$\hbox{V}_{\hbox{\scriptsize In}}$ }

\def\VP      {$\hbox{V}_{\hbox{\scriptsize P}}$}
\def\VPs     {$\hbox{V}_{\hbox{\scriptsize P}}$ }

\def\PI      {$\hbox{P}_{\hbox{\scriptsize In}}$}
\def\PIs     {$\hbox{P}_{\hbox{\scriptsize In}}$ }

\def\IP      {$\hbox{In}_{\hbox{\scriptsize P}}$} 
\def\IPs     {$\hbox{In}_{\hbox{\scriptsize P}}$ }

\def\PiP     {$\hbox{P}_{\hbox{\scriptsize i(P)}}$}
\def\PiI     {$\hbox{P}_{\hbox{\scriptsize i(In)}}$}
\def\PiH     {$\hbox{P}_{\hbox{\scriptsize i(HEX)}}$}
\def\PiPs    {$\hbox{P}_{\hbox{\scriptsize i(P)}}$ }
\def\PiIs    {$\hbox{P}_{\hbox{\scriptsize i(In)}}$ }
\def\PiHs    {$\hbox{P}_{\hbox{\scriptsize i(HEX)}}$ }

\def\IiP     {$\hbox{In}_{\hbox{\scriptsize i(P)}}$}
\def\IiI     {$\hbox{In}_{\hbox{\scriptsize i(In)}}$}
\def\IiH     {$\hbox{In}_{\hbox{\scriptsize i(HEX)}}$}
\def\IiPs    {$\hbox{In}_{\hbox{\scriptsize i(P)}}$ }
\def\IiIs    {$\hbox{In}_{\hbox{\scriptsize i(In)}}$ }
\def\IiHs    {$\hbox{In}_{\hbox{\scriptsize i(HEX)}}$ }

\def\zvps 	{$\left[\hbox{Zn}_{\hbox{\scriptsize In}}\hbox{-}\hbox{V}_{\hbox{\scriptsize P}}\right]$ }

\begin{document}

\title{Finite size scaling as a cure for supercell approximation errors in calculations of neutral native defects in InP.}

\author{C W M Castleton$^{1,2}$ and S Mirbt$^{3}$}

\address{1 Material Physics, Materials and Semiconductor Physics Laboratory, 
Royal Institute of Technology (KTH), Electrum 229, 16440 Kista, 
Sweden.} 
\address{ 2 Department of Physical Electronics/Photonics, 
ITM, Mid Sweden University, 85170 Sundsvall, Sweden.} 
\address{ 3 Theory of Condensed Matter, Department of Physics, Uppsala University, 
Box 530, 75121 Uppsala, Sweden.}

\date{\today}

\begin{abstract}
The relaxed and unrelaxed formation energies of neutral antisites and 
interstitial defects in InP are calculated using ab initio density 
functional theory and simple cubic supercells of up to 512 atoms.  The finite size errors in 
the formation energies of all the neutral defects arising from the 
supercell approximation are examined and corrected for using finite 
size scaling methods, which are shown to be a very promising approach 
to the problem.  Elastic errors scale linearly, whilst the errors 
arising from charge multipole interactions between the defect and its 
images in the periodic boundary conditions have a linear plus a higher 
order term, for which a cubic provides the best fit.  These latter errors are 
shown to be significant even for neutral defects.  Instances are also 
presented where even the 512 atom supercell is not sufficiently 
converged. Instead, physically relevant results can be obtained only by finite size scaling the
results of calculations in several supercells, up to and including the 512 atom cell and in extreme cases possibly even including the 1000 atom supercell. 
\end{abstract}
\maketitle

\section{Introduction}
\label{Intro}

Over the past decade or so first principles density functional theory 
(DFT) \cite{DFT} has become a powerful tool for studying the 
properties of defects in semiconductors.  It is possible to calculate 
formation energies, binding energies and migration barriers, to 
predict local structure and, up to a certain point, defect levels and 
electrical activity.  Problems and limitations remain, however, and 
one of the greatest is the very limited size of the systems for which 
calculations are feasible: 10s or 100s of atoms, even when we wish to 
describe physical problems involving 1,000s or 10,000s.  This leaves 
the results heavily influenced by errors arising from the boundary 
conditions.  These errors must therefore be carefully studied so that 
their effects can be understood and accounted for when results are 
interpreted.  There are two types of boundary conditions commonly 
used: open and periodic.  Open boundary conditions are usually 
encountered in cluster calculations.  The surface atoms are 
``terminated'' with hydrogen to use up spare electrons, but are 
otherwise surrounded by empty space.  Periodic boundary conditions 
(PBCs), meanwhile, are found in supercell calculations, in which a 
block of atoms is surrounded not by empty space but by an infinite 
array of copies of itself.  Both approaches have strengths and 
weaknesses. We present here  a detailed study of the problems arising from the use 
of the supercell approximation and we propose a method to overcome them.

Finite size errors in supercell calculations come from two main sources.  Elastic 
errors often arise because the supercell is simply not large enough to 
contain all of the local relaxation around the defect, leaving  
calculated formation energies too high.  In addition, the defect 
interacts with an infinite array of spurious ÒimagesÓ of itself seen 
in the PBCs, via both elastic and electrostatic interactions.  The 
direct elastic interactions can easily be truncated by 
freezing all atoms lying on 
the surface of the cell at their ideal lattice positions.  The 
electrostatic interactions, on the other hand, cannot be truncated or 
removed.  They result in errors in the calculated formation, binding 
and migration energies, errors which can be on the same order as the 
energies themselves.  For practical supercell sizes they need not even 
be negligible for neutral defects, since dipolar and quadrupolar 
interactions can remain significant.  These latter can even result in 
errors in the calculated structures, since they favour certain 
symmetries and local relaxation modes over others.  Hence 
indirect elastic errors cannot be avoided either. Finally, a third source of finite size errors is also present: the defect state wavefunctions can overlap with their images in the PBCs leading to a spurious dispersion of the defect 
levels which in turn can affect the formation energies, especially if the defect level is only partially filled. 
The errors related to this dispersion (or tunnelling) are expected to have only a fairly small and rather short ranged (exponential) effect.

Recently, various correction schemes have been suggested 
\cite{MP,OtherKorr} to compensate for at least the leading terms in 
the errors arising from the electrostatic interactions.  They are 
usually based upon fits to quasi-classical models and/or multipole 
expansions of the electrostatic interactions.  They have met with 
varying levels of success but, so far at least, are generally 
considered insufficiently reliable.  There are more direct approaches, 
however.  Probert and Payne \cite{Matt} recently presented a detailed 
ab initio study of the neutral vacancy in Si, considering all aspects 
of convergence, from basis set and k-point sampling to size and 
symmetry of supercells.  They demonstrated that the use of ``large'' 
supercells (200+ atoms) can be essential for obtaining the correct 
physical results.  Meanwhile, we recently presented \cite{NeutralV} a 
study of the neutral vacancies in InP.  We demonstrated the advantages 
of not only using large supercells but also finite size scaling the 
results obtained.  We evaluated both the relaxed and unrelaxed 
formation energies of the phosphorus (\VP) and indium (\VI) vacancies 
in simple cubic supercells of 8, 64, 216 and 512 atoms.  We then 
showed that the variation in the formation energy with supercell size, 
$L$, follows rather closely the form
\begin{equation}\label{scale_eqn}
E_{\rmd}^{\rmC}\!(L) = E_{\rmd}^{\infty} + a_{1}L^{-1} + a_{3}L^{-3}
\end{equation}
\noindent where $E_{\rmd}^{\rmC}\!(L)$ is the formation energy in 
supercell ``$\rmC$'' and $a_{1}$ and $a_{3}$ are fitting parameters.  
$E_{\rmd}^{\infty}$ is the finite size scaled formation energy 
corresponding to an infinitely large supercell.  Equation 
\ref{scale_eqn} has, in fact, the same form  
($L^{^-1}$ plus $L^{^-3}$) as the corrections proposed by Makov and 
Payne \cite{MP}.  We will return to the issue of whether or 
not this is the correct form in section \ref{Scaling form}.  For the 
vacancies we showed that the error bars on the values obtained for 
$E_{\rmd}^{\infty}$ are usually rather small: $\Or$(0.1 eV) or less, 
depending also upon the level of k-point convergence in individual 
cells.  Care must be taken, however, when scaling relaxed formation 
energies of strongly Jahn-Teller \cite{Jahn-Teller} active defects, 
such as \VP.  In such cases rather wider error bars are obtained if 
the symmetry of the relaxed structures varies with supercell size.  
Numerically, we also found that (for example) $E_{\rmd}^{\infty}$ for 
the unrelaxed \VIs is actually $\sim$0.2 eV above the value obtained 
in the 512 atom supercell, suggesting that there are cases 
for which even the 512 atom cell is not large enough to be converged, 
so that scaling becomes essential.  

In the present paper we extend the study to that of the other neutral 
native defects: the antisites and interstitials. It should be noted, however, that our primary purpose is not the study of the defects themselves but of the finite size errors which arise when calculating their formation energies. InP has the 
zinc-blend structure, with two antisites, \PIs and \IP, both with 
tetragonal (T$_{\hbox{\scriptsize d}}$) symmetry when unrelaxed.  For 
interstitials there are three high symmetry sites: two tetragonal, 
with four In or four P nearest neighbours (X$_{\hbox{\scriptsize 
i(In)}}$ or X$_{\hbox{\scriptsize i(P)}}$) and a quasi-hexagonal site 
(X$_{\hbox{\scriptsize{i(hex)}}}$) with six nearest neighbours (three 
P and three In) and C$_{\hbox{\scriptsize 3v}}$ symmetry.  Previous 
work for the isolated neutral cases has mostly been limited to the 64 
atom supercell.  Here, formation energies were around 5-6 eV for the 
tetrahedral interstitials \cite{Old interstitials} and around 3 eV for 
the antisites \cite{Old antisites}.  \IPs displayed some Jahn-Teller 
behaviour whilst \PIs did not.  We will examine how these results 
change with larger supercells and with scaling.  In the next section 
we describe the method to be used.  In section \ref{Scaling results} 
we will describe the basic scaling results for the various neutral 
defects and will examine in sections \ref{Scaling form} and \ref{scale 
problems} the form of the scaling and when it does and does not work.  
In section \ref{Non-neutral} we consider briefly other charge states 
of the defects.  In Section \ref{Linear} we discuss the origin of the 
surprising linear scaling we find in certain cases. In 
section \ref{Other errors}~we estimate the size of the other non-finite size dependent errors for the purpose of comparison, before concluding in \ref{Conclusions}.

\section{Method and k-point convergence.}
\label{Method}

\begin{figure}
\epsfxsize=6.0cm
\epsfbox[108 28 638 574]{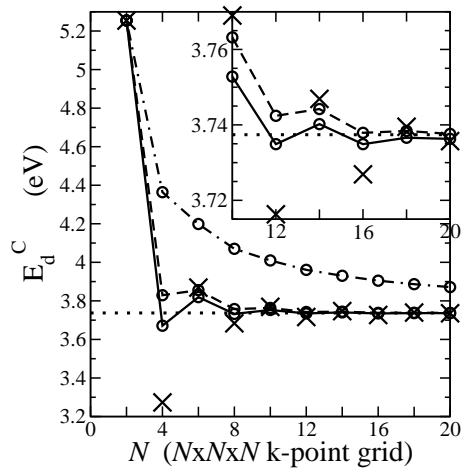} \caption{\label{Kconv} Convergence of formation 
energy $E_{d}^{C}\left(N\right)$ with even $N$x$N$x$N$ Monkhorst-Pack 
k-point grids.  $\times$: calculated values of formation energies.  
Dotted line: ``converged'' value (calculated using 30x30x30 k-point 
grid).  Dot-dashed: running average over 
$E_{\rmd}^{\rmC}\!\left(N\right)$.  Dashed line: running average 
weighted by the number of k-points in the irreducible Brillouin zone.  
Solid line: running average weighted by the total number of k-points 
in the full Brillouin zone.}
\end{figure}

We use planewave ab initio DFT \cite{DFT} within the local density 
approximation (LDA) together with ultrasoft pseudopotentials
\cite{USPP} (US-PP) using the VASP code \cite{VASP}.  We recently presented 
\cite{ZVP} a study of the \zvps complex in InP using the same 
technique and potentials.  Unfortunately, physical memory limitations in even the largest parallel computer 
facilities available to us mean that we must treat the 4d electrons of In as core, even though they are comparatively 
shallow. (A calculation for bulk InP in its FCC primitive cell places the In 4d states about 14.5 eV below the valence 
band edge at the $\Gamma$ point when they are treated as valence - fairly deep but still close enough to potentially make 
some contributions to formation energies.)
It would be preferable to treat them as valence, but then properly k-point converged calculations in the 512 
atom cell - which our analysis requires - would be impossible. The size of the resulting error will be 
examined in section \ref{Other errors}.

The optimized LDA lattice constant using these pseudopotentials is 
5.827 \AA~and the band gap is 0.667 eV, compared to 5.869 \AA~ and 1.344 
eV in experiment.  As stated above, we will only use simple cubic 
supercells of 8, 64, 216 and 512 atoms, since it is important to keep to 
a single supercell symmetry since the scaling is different for different 
symmetries. No restrictions are placed upon the symmetry of 
relaxations, but we do not allow atoms located on the surface of the 
cell to relax.  The exception is interstitials at the quasi-hexagonal 
sites in the 8 atom cell.  Here, three of the nearest neighbours lie 
on the surface.  Initial tests showed that the relaxation was not 
stable if all of these were allowed to relax at once so relaxations 
were done in stages: firstly one group of neighbours was relaxed 
whilst the others were kept fixed, then vice verse.

The key quantity is the formation energy
\begin{equation}
E_{\rmd}^{\rmC} = E_{\rmT}^{\rmC}\!\left(\hbox{defect}\right) - 
E_{\rmT}^{\rmC}\!\left(\hbox{bulk}\right) + \sum_i \mu_in_i
\end{equation}
\noindent where $E_{\rmT}^{\rmC}\!\left(\hbox{defect}\right)$ and 
$E_{\rmT}^{\rmC}\!\left(\hbox{bulk}\right)$ are the total energy of the 
supercell with and without the defect, calculated using the same values 
of planewave cutoff, k-point grid, etc, to make use of the cancellation 
of errors.  The defect is formed by adding/removing $n_i$ atoms of 
chemical potential $\mu_i$.  We use $\mu_{\hbox{\scriptsize P}}$ = 
3.485 eV and $\mu_{\hbox{\scriptsize In}}$ = 6.243 eV, \cite{ZVP} 
corresponding to stoiciometric conditions.  A planewave cutoff energy 
of 200 eV and a Monkhorst-Pack 4x4x4 k-point grid 
\cite{Monkhorst-Pack} was found sufficient \cite{ZVP} for converged 
non-relaxed calculations in the 64 atom supercell with errors around 
$\Or$(0.01 eV).  Hence grids of 8x8x8 in the 8 atom cell and 2x2x2 in 
the larger cells should suffice.  However, in this study we examine 
specifically the errors arising from the supercell approximation 
itself. K-point convergence is different in different supercells and we do not wish to include any significant errors due 
to this. Hence we need to ensure even higher levels of convergence during this study - much higher than is normally 
required or practical. (Examining the scaled properties of the native InP defects themselves, 
over all relevant charge states, is left to future work, however, as it will be done with rather different 
levels of k-point convergence, pseudopotentials etc, once we have in the present work established the behaviour of the finite size errors.)
 We use k-point 
grids of up to and including 12x12x12, 8x8x8, 4x4x4 and 4x4x4 in the 
8, 64 216 and 512 atom cells respectively.  (Exceptions are: \PI - 
only 2x2x2 needed in the 512 atom cell, and \IiIs and \IiPs - 6x6x6 
used in the 216 atom cell to check convergence.) To improve 
convergence further we use weighted averages over $E_{\rmd}^{\rmC}$ 
values calculated using a series of grids.  Fig \ref{Kconv} shows the 
advantages clearly: the unrelaxed formation energies 
$E_{\rmd}^{008}\!\left(N\right)$ for \PiIs in the 8 atom cell are 
shown, calculated using $N$x$N$x$N$ k-point grids and plotted against 
$N$.  (This case has the most difficult k-point convergence in this 
paper.) To get errors safely below $\Or$(0.005 eV) a k-point grid of 
at least 18x18x18 is needed.  Taking the average over all the 
$E_{\rmd}^{\rmC}$ values up to a particular $N$x$N$x$N$ (dot-dashed line) 
is unhelpful, but taking a weighted average
\begin{equation}
{\overline{E_{\rmd}^{\rmC}}} = \frac{\sum_{N} w_{N} E_{d}^{C}\!\left(N\right)}{\sum_{N} w_{N}}
\end{equation}
\noindent helps dramatically, as it effectively increases the k-point 
density: the points in a 4x4x4 grid are not contained in a 6x6x6 
grid, for example.  There are two obvious choices for $w_{N}$: the 
best is $w_{N}$ = $N^3$, the number of k-points in the full Brillouin 
zone, but setting $w_{N}$ equal to the number in the irreducible wedge 
is not bad.  (Solid and dashed lines respectively.) All subsequent 
results will be weighted averages using $w_{N}$ = $N^3$.  
(Incidentally, for the unrelaxed neutral vacancies we find 
\cite{NeutralV} errors of 0.36 eV and 0.06 eV respectively for \VPs 
and \VIs in the 512 atom cell when comparing $\Gamma$ point only 
calculations to converged values, so the $\Gamma$ point is never 
sufficient.)

In reference \cite{ZVP} we also showed that the relaxation energy
\begin{equation}
\epsilon_{\hbox{\scriptsize Relax}}\!\left(N\right) = 
E_{\rmd}^{\rmC\rmRx}\!\left(N\right) - 
E_{\rmd}^{\rmC\rmId}\!\left(N\right)
\end{equation}
\noindent (where $E_{\rmd}^{\rmC\rmRx}\!\left(N\right)$ and 
$E_{\rmd}^{\rmC\rmId}\!\left(N\right)$ are 
$E_{\rmd}^{\rmC}\!\left(N\right)$ with atoms at relaxed and ideal 
positions respectively) converges faster with k-point grid than 
$E_{\rmd}^{\rmC\rmRx}\!\left(N\right)$ and 
$E_{\rmd}^{\rmC\rmId}\!\left(N\right)$ themselves. The relaxed formation energies 
$E_{\rmd}^{\rmC\rmRx}$ are then approximated by
\begin{equation}
{\overline{E_{\rmd}^{\rmC\rmRx}}} \approx 
{\overline{E_{\rmd}^{\rmC\rmId}}} -
\epsilon_{\hbox{\scriptsize Relax}}\!\left(N\right)
 			   = {\overline{E_{\rmd}^{\rmC\rmId}}} + 
E_{\rmd}^{C\rmRx}\!\left(N\right) - E_{\rmd}^{\rmC\rmId}\!\left(N\right)
\end{equation}
\noindent The relaxation energies used will be weighted averages using 
6x6x6 and 8x8x8 Monkhorst-Pack k-point grids in the 8 atom cell, 2x2x2 
and (if the convergence is uncertain) 4x4x4 grids in the 64 atom cell 
and 2x2x2 in the 216 and 512 atom supercells.  For the latter two we 
usually restrict the k-point grid to the irreducible Brillouin zone of 
the undisturbed bulk lattice.  In other words, we use just the special 
k-point (0.25,0.25,0.25).  This amounts to assuming that the distortion 
in the bandstructure due to the presence of the defect is either 
localized (thus important only very near $\Gamma$) or symmetric.  It 
introduces a small error whose significance disappears in the large 
supercell limit.

\begin{figure}
\epsfxsize 8.5 cm
\epsfbox[15 26 817 568]{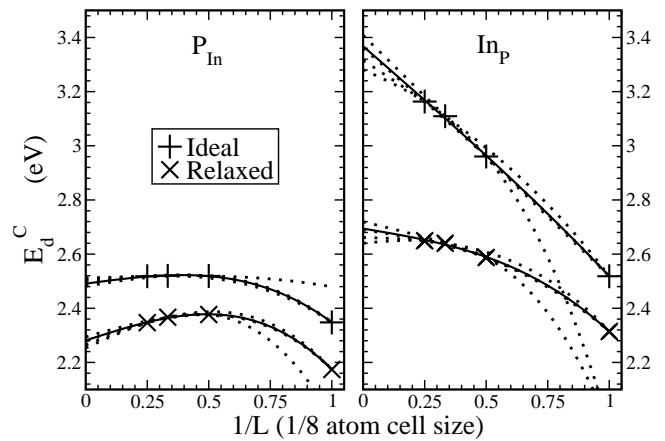} \caption{\label{Scale_A} Scaling the formation 
energy for the relaxed and unrelaxed structures of the neutral 
antisite defects \PIs and \IPs with inverse supercell size.  The 
$x$-axis scale is in units of 1/(5.827\AA), the inverse of the 8 atom 
supercell size.  Hence $x = $1.00, 0.50, 0.33 and 0.25 correspond to 
the 8, 64, 216 and 512 atom cells respectively.  }
\end{figure}

\section{Scaling results.}
\label{Scaling results}

\begin{figure}
\epsfxsize 8.5truecm
\epsfbox[15 26 816 568]{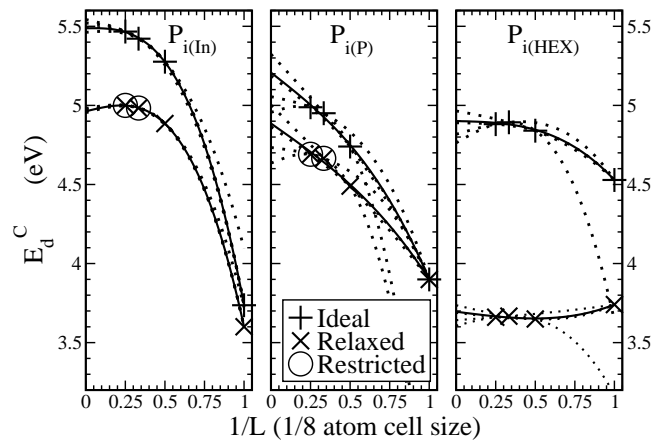}
\caption{\label{Scale_Pi} 
Scaling the formation energy of neutral phosphorus interstitials.  
Those at the tetrahedral sites \PiIs and \PiPs are not stable in the 
216 and 512 atom supercells: if the symmetry is not restricted then the 
interstitials relax towards the hexagonal site.  Plotted points 
are from relaxations in which the symmetry was restricted to T$_{\rmd}$ 
to prevent this.}
\end{figure}

\begin{figure}
\epsfxsize 8.5truecm
\epsfbox[15 26 816 568]{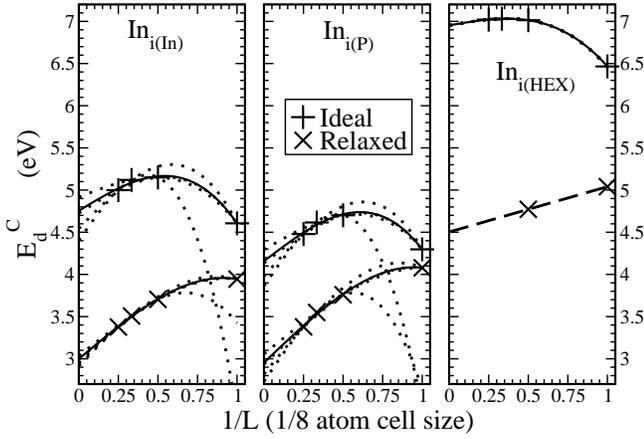}
 \caption{\label{Scale_Ini} 
Scaling the formation energy of neutral indium interstitials.  Indium 
is not stable at the hexagonal site \IiHs in the 216 and 512 atom 
supercells and cannot be forced to stay put by restricting relaxation 
symmetry.}
\end{figure}

In Figs \ref{Scale_A}, \ref{Scale_Pi} and \ref{Scale_Ini} we show the 
formation energies for the antisites, phosphorus interstitials and 
indium interstitials respectively, both relaxed (minimum energy 
configurations) and unrelaxed (atoms at ideal bulk lattice sites) 
plotted against inverse supercell size.  The solid lines are fits of the 
four points to equation \ref{scale_eqn}.  The $y$-axis intersect of each of 
these fits is the scaled formation energy $E_{\rmd}^{\infty}$ 
corresponding to the formation energy of a single isolated defect in 
an otherwise perfect, infinite lattice.  The inclusion of formation 
energies from the 8 atom supercell could be questioned, since in
itself it is so far from being converged.  However, the results shown in 
the figures clearly support our expectation that the correct form for 
the scaling has (at least) three parameters.  We therefore need 
results from at least four supercells of the same symmetry.  It would 
have been preferable to use the 1000 atom simple cubic cell, but 
sufficiently converged ab initio calculations for InP defects in this 
cell are not possible with current facilities.  In addition, the 
results themselves justify the use of the 8 atom cell: the scaling 
mostly works very well, producing small error bars and with 
$E_{\rmd}^{008}$ lying on or near the curves.  This, incidentally, 
also tells us that the k-point convergence in the individual cells was 
sufficient.  The cases where the scaling does not work so well turn 
out to be due to other problems, see section 
\ref{scale problems}.

To get an idea of the accuracy of the 
fitting and of the derived $E_{\rmd}^{\infty}$ values (and the 
individual $E_{\rmd}^{\rmC}$) four more fits (dotted lines) 
are added in each case.  For each of these, one of the four data points 
has been omitted.  The spread in the resulting $y$-axis intersects 
gives an error bar for the scaled formation energy 
$E_{\rmd}^{\infty}$.  (It should be emphasized that the dotted lines - 
three parameters to three points - are not intended to be meaningful 
in themselves but are merely indications of the scale of the error in 
the real solid line fits.) The scaled formation energies are listed in 
table \ref{Final Ed}, together with those of the vacancies taken from 
Ref [5
].

The scaling curves also have a more general and rather practical meaning: they are essentially predictions of the 
formation energies in {\it all }Êsimple cubic supercells from 8 atoms to infinity. For example, table \ref{Scale predict} 
shows the predicted formation energies in the 1000 and 8000 atom supercells, as they would arise 
from k-point and basis set converged LDA calculations. (The 64,000 atom cell would be just as easy, but these predictions 
are more likely to be tested and - hopefully - confirmed within 
our lifetimes!) As a more immediate test, table \ref{Scale predict} also shows the formation energies in the 512 atom 
cell predicted by scaling the results from only the 8, 64 and 216 atom cells: following along that dotted line which doesn't pass exactly through the 512 atom value for each case in the figures. The error in this prediction (as compared to the actual 
calculated values) is also shown. The errors are pleasantly small, especially considering only 3 cells have been used, including the 8 atom one. The errors in the other predictions are expected to be much smaller still.

\begin{figure}
\epsfxsize 8.5truecm
\epsfbox[25 26 811 568]{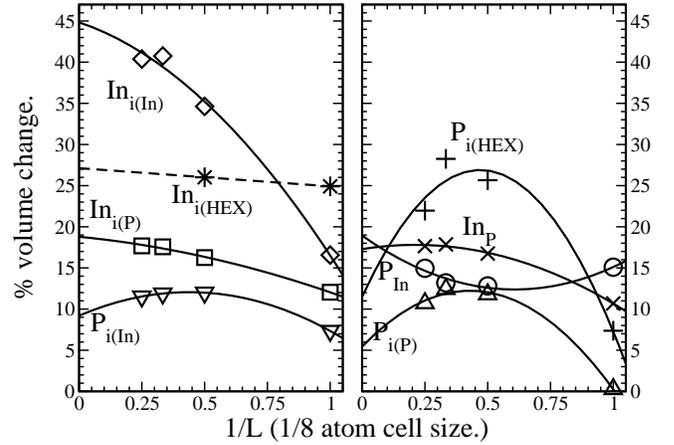}
\caption{\label{Volume} 
Scaling of the volume change with supercell size: percentage volume 
change upon relaxation is plotted against inverse supercell size.  
\PIs and \PiPs relax inwards, the others outwards.}
\end{figure}

All the relaxed structures turn out to be symmetric, with just breathing mode
relaxations.  The exception is \IP, which shows some moderate 
Jahn-Teller behaviour (see section \ref{JT effects}. )
All the interstitials apart from \PiPs relax outwards, as does \IP, 
whilst \PIs relaxes inwards.  This is all as expected, since P is 
smaller than In.  Fig \ref{Volume} shows the scaling of the 
percentage volume change upon relaxation.  (The volume shown is that 
of the polyhedron defined by the nearest neighbours.) The fits are 
described in section \ref{Scaling form} and the scaled results are in 
table \ref{Final Ed}.  The error bars are derived in the same way as 
for the formation energies, although the dotted lines are omitted here 
for clarity.

The most stable neutral native defect in stoiciometric InP turns out to 
be the phosphorus antisite \PIs closely followed by the  
vacancy \VPs and then the indium antisite \IPs.  Then come the 
interstitials, of which indium is the more stable. The least stable 
neutral defect is \VI. The most stable site for the phosphorus interstitial is the 
quasi-hexagonal site \PiH.  For indium, the two tetrahedral sites are 
degenerate to within the error bars but numerically the phosphorus 
surrounded site \IiPs is 0.05 eV lower - which is probably  
correct, since the error bars are asymmetric.  (See 
Fig \ref{Scale_Ini}.) This is a result which is only apparent in very 
large supercells and the degeneracy only appears at the 512 atom 
supercell.  In smaller cells \IiIs seems more stable.  What is more, 
scaling shows that simply taking the 512 atom result would give a 
relaxed formation energy about 0.4 eV too high (0.8 eV too high if we 
stopped at 64 atoms) which can be large enough to make real 
differences in the predicted physics of InP.  As with the formation 
energy \cite{NeutralV} of \VIs this emphasizes the value of finite size scaling the 
results of supercell calculations, since the largest cells for which we can actually do 
calculations can still be too small to be fully converged.

There are complications with the stable interstitial sites as a 
function of supercell size.  Whilst both interstitials are stable at 
all three locations in the two smallest cells they are not so in the 
larger cells.  For indium the quasi-hexagonal site lies about 
$\sim1\frac{1}{2}$ eV above the tetrahedral sites and is not stable in 
the larger cells.  An indium atom placed here migrates to a 
tetrahedral site upon relaxation, indicating that the quasi-hexagonal 
site is probably not even metastable for indium in the real material.  
For the phosphorus interstitial, on the other hand, it is the 
tetrahedral sites which are unstable.  In this case, however, we can 
still obtain (hypothetical) relaxed formation energies at the 
tetrahedral sites by only allowing T$_{\rmd}$ breathing mode 
relaxations.  The tetrahedral sites are 1.2 eV higher than the 
quasi-hexagonal site.  The reason for this difference in stabilities 
may be (partially) stearic: the unrelaxed nearest neighbour 
distances are shorter at the hexagonal site, where the smaller 
phosphorus interstitials are stable and longer at the tetrahedral 
site where the larger indium interstitial sits.

\begin{table}
\caption{\label{Final Ed}Scaled relaxed and unrelaxed (ideal 
lattice sites) formation energies $E_{\rmd}^{\infty}$, plus the scaled 
percentage volume change (\%$\delta$V) upon relaxation for neutral 
native defects in InP, listed in order of (relaxed) stability in 
stoiciometric material.  (The volume is that of the polyhedron defined 
by the nearest neighbours.) Note that the error bars are not actually 
symmetric: see Figs \ref{Scale_A} to \ref{Scale_Ini}.}
\begin{center}
\begin{tabular}{llll}
\hline\hline
Defect & Ideal (eV)              & Relaxed (eV)             & \%$\delta$V       \\\hline
\PI    & 2.49$\pm$0.02           & 2.28$\pm$0.03            &-19$\pm$7          \\
\VP    & 3.00$\pm$0.10$^{\rm c}$ & 2.35$\pm$0.15$^{\rm c}$  &-43$\pm$4$^{\rm c}$\\
\IP    & 3.37$\pm$0.09           & 2.69$\pm$0.06            & 17$\pm$3          \\
\IiP   & 4.15$\pm$0.30           & 2.95$\pm$0.20            & 19$\pm$3          \\
\PiH   & 4.90$\pm$0.14           & 3.69$\pm$0.08            & 12$\pm$38         \\
\PiP   & 5.21$\pm$0.16           & 4.88$\pm$0.25$^{\rm a}$  &-5$\pm$7$^{\rm a}$ \\
\PiI   & 5.49$\pm$0.06           & 4.96$\pm$0.02$^{\rm a}$  & 9$\pm$1$^{\rm a}$\\
\IiI   & 4.75$\pm$0.35           & 3.00$\pm$0.08            & 45$\pm$19         \\
\IiH   & 6.95$\pm$0.01           & $\sim3.5^{\rm b}$        & $\sim27^{\rm b}$ \\
\VI   & 4.95$\pm$0.10$^{\rm c}$  & 4.20$\pm$0.05$^{\rm c}$  &-35$\pm$3$^{\rm c}$\\\hline\hline
\end{tabular} 
\nl
$^{\rm a}$ Unstable in some cells, value results from symmetry 
restricted relaxations, see text.\nl $^{\rm b}$ Unstable in some 
cells, value is rough extension with no error bar available, see 
text.\nl $^{\rm c}$ Data taken from Ref. [5
] Scaling re-done using $n=2$.\nl
\end{center}
\end{table}

\begin{table}
\caption{\label{Scale predict}ÊPredictions for the unrelaxed and relaxed formation energies in the (to date) uncalculated 8000 and 1000 atom simple cubic supercells. Also shown are predictions for the 512 atom supercell formation energies 
obtained by scaling the values in the 8, 64 and 216 atom cells, together with the difference (Error) between the predicated and calculated values. (Energies in eV.)}
\begin{center}
\begin{tabular}{llllllllllllll}
\hline\hline
       &\multicolumn{6}{c}{Idea Structures} &         &\multicolumn{6}{c}{Relaxed Structures}           \\\hline
Defect &8000 &\quad  &1000 &\quad  &512   &Error   &\quad\quad  &8000 &\quad  &1000 &\quad  &512    &Error  \\\hline
\PI    &2.50 &       &2.51 &       &2.51  &-0.01   &            &2.31 &       &2.33 &       &2.35   &0.00   \\
\VP    &2.99 &       &2.98 &       &3.00  &0.04    &            &2.40 &       &2.46 &       &2.46   &-0.03  \\
\IP    &3.29 &       &3.21 &       &3.18  &0.02    &            &2.68 &       &2.66 &       &2.66   &0.01   \\
\IiP   &4.30 &       &4.44 &       &4.55  &0.07    &            &3.13 &       &3.31 &       &3.43   &0.05   \\
\PiH   &4.90 &       &4.89 &       &4.91  &0.03    &            &3.68 &       &3.67 &       &3.68   &0.02   \\
\PiP   &5.13 &       &5.05 &       &5.05  &0.06    &            &4.81 &       &4.74 &       &4.75   &0.06   \\
\PiI   &5.49 &       &5.48 &       &5.45  &0.02    &            &4.99 &       &5.00 &       &5.00   &0.00   \\
\IiI   &4.87 &       &4.97 &       &5.08  &0.08    &            &3.15 &       &3.31 &       &3.40   &0.02   \\
\IiH   &6.98 &       &7.01 &       &7.02  &0.00    &            & -   & -     & -   &       & -     & -     \\
\VI    &4.88 &       &4.80 &       &4.77  &0.02    &            &4.19 &       &4.18 &       &4.17   &-0.02  \\\hline\hline
\end{tabular} 
\nl
\end{center}
\end{table}

\section{The correct form for the scaling equation.}
\label{Scaling form}

\subsection{Formation and relaxation energies.}

So far it has been assumed that the correct 
functional form for scaling is that of equation \ref{scale_eqn}.  This 
need not be the case, however.  Equation \ref{scale_eqn} is based upon 
approximations and predictions for the form of the errors arising from 
electrostatic charge multipole interactions between the defect and 
it's images in the PBCs and these could be incorrect.  In addition, 
we include here relaxations within finite sized supercells, so we have 
both elastic errors and, potentially, cross terms between the elastic 
and electrostatic errors, so other possible scalings should be 
considered.  There is clearly a linear term present, plus at least one 
higher order term, so we consider scaling of the form:
\begin{equation}\label{nscaling} 
	E_{\rmd}^{\rmC}\!(L) = E_{\rmd}^{\infty} + a_{1}L^{-1} + a_{n}L^{-n}
\end{equation}
\noindent with $n=2$, 3 and 4.  The fit quality is assessed 
using the ``$\chi^2$'' test.  This is not easy to do reliably with 
only four points, so we average over different defects. 
Simply summing or averaging the $\chi^2$ values would bias the 
conclusion towards the worst data sets.  Instead, for each data set we 
find $\chi^2$ for each value of $n$, select the $n$ giving the best fit 
($\chi^2_{\hbox{best}}$) and then calculate a quality factor 
$Q_{n}$=$\chi^2_{n}$/$\chi^2_{\hbox{best}}$ for each $n$.  
We then compare the averaged $\overline{Q_n}$.

\begin{figure}
\epsfxsize 8.5truecm
\epsfbox[21 26 811 575]{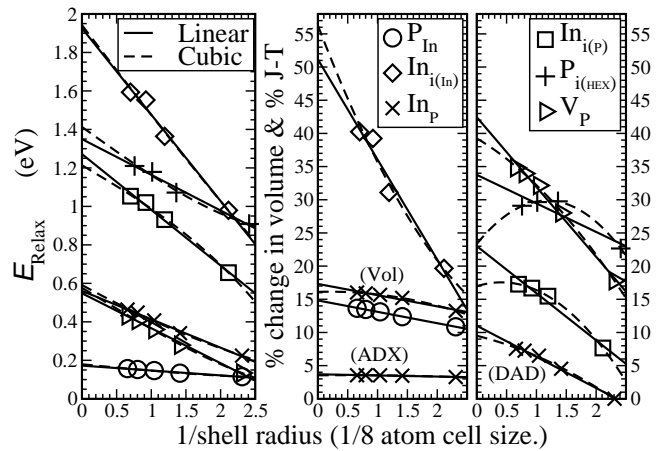}
\caption{\label{Shells} 
The relaxation energy ($\epsilon_{\hbox{\scriptsize Relax}}$) plus the \% 
volume change as a function of 
the number of shells around a defect which are allowed to relax.  For \IP the \% Jahn-Teller distortion in the ADX and DAD structures (see main text) is also shown. (Relaxation energy and volume change are only shown for the DAD structure: those for the ADX structure are very similar.) The 
216 atom supercell is used, giving 1$\rightarrow$5 shells for 
vacancies and antisites, 1$\rightarrow$4 for interstitials.  The 
$x$-axis is the inverse of the radius of the atom shell, in units of one 
over the 8 atom cell size.  Volume changes are outwards for all except 
\PI.  Fits are to equation \ref{nscaling} with (dashed) and 
without (solid) the $L^{-n}$ term.  ($n=2$ for $\epsilon_{\hbox{\scriptsize Relax}}$,
$n=3$ for the volumes.) Results for \VPs are taken from 
Ref. \cite{NeutralV}.}
\end{figure}

We first examine the scaling of the elastic errors.  To do this we 
have performed a series of relaxations in the 
216 atom cell for each of six defects.  In these relaxations, the 
number of shells of atoms permitted to relax around the defect is 
varied from 1$\rightarrow$4 for interstitials or 1$\rightarrow$5 for 
antisites and vacancies.  (4 and 5 are the maximum numbers of 
complete shells which fit inside the supercell.) Since the cell size 
is kept constant the electrostatic interactions will be (almost) 
constant so any variation in $E_{\rmd}$ is due to elastic effects.  
In the left panel of Fig \ref{Shells} we show the scaling of 
the relaxation energy with the inverse radius of the outermost 
relaxing shell.  (The units have been scaled to match those in 
previous figures.) The scaling is almost purely linear (solid lines in 
the figure) even - within the bounds of error - for the Jahn-Teller 
active defects like \IPs and \VP.  Adding a higher order term, such as 
$L^{-3}$ (dashed lines) clearly only improves the fit very slightly: 
$\chi^2$ is reduced by about 30\% on average.

Having established that the scaling of the relaxation energy, 
and hence of the elastic errors in $E_{\rmd}^{\infty}$, is linear 
we turn to the scaling of the formation energies with 
supercell size.  Unfortunately there is too much scatter in several of 
the curves so that whilst $n=3$ is best in some cases, in others it comes 
an often close second to $n=2$ or $n=4$.  However, since the elastic 
errors are now known to be linear in relaxation radius and hence in 
supercell size, we can assume that it will simply add to the linear 
term in equation \ref{nscaling}, so that we can calculate 
$\overline{Q_n}$ using all 19 data sets (Figs \ref{Scale_A} to 
\ref{Scale_Ini} and Ref. [5
]).  The result obtained is that 
the scaling does indeed fit best with $n=3$, with both 
$\overline{Q_2}$ and $\overline{Q_4}$ being 4 times larger than 
$\overline{Q_3}$.  In fact, if we calculate the $\{\overline{Q_n}\}$ 
using only the unrelaxed formation energies we still find $n=3$ fits 
best and if we use the relaxed formation energies minus that of \PIs 
we find the same result.  Unfortunately if we use all of the relaxed 
energies including \PIs we get a different result: $n=2$ provides a 
better fit.  This suggests the (faint) possibility of a cubic scaling 
for unrelaxed formation energies but quadratic for relaxed energies, 
which shifts the predicted $E_{\rmd}^{\infty}$ by 0.01$\rightarrow$0.2 
eV depending on the defect.  A quadratic scaling would seem odd, since 
it comes from neither the elastic nor the electrostatic errors but it 
could possibly arise from the cross terms between them. An alternative and perhaps more likely explanation is that the uncertainty regarding $n$
is a result of the spurious dispersion of the defect levels mentioned above. This adds an exponential term to the formation energies in the smallest 
supercells, blurring the picture slightly. It seems most likely that the scaling {\it should} be cubic even for the relaxed formation 
energies: even now it is only one defect which seems to particularly disagree.  This should be confirmed once reliable 
calculations involving the 1000 atom supercell become feasible.

\begin{table}
\caption{\label{Vol table} Scaled percentage volume changes upon 
relaxation, from scaling with number of shells relaxed in the 216 
atom cell and from scaling with supercell size.}
\begin{center}
\begin{tabular}{lll}
\hline\hline
Defect & Shells    & Superells \\\hline
\PI    &-15$\pm$1  &-19$\pm$7  \\
\IP    & 20$\pm$2  & 17$\pm$3  \\
\IiP   & 17$\pm$1  & 19$\pm$3  \\
\IiI   & 51$\pm$40 & 45$\pm$19 \\
\PiH   & 23$\pm$2  & 12$\pm$38 \\
\VP    &-39$\pm$3  &-43$\pm$4  \\\hline\hline
\end{tabular} 
\end{center}
\end{table}

\subsection{Relaxed volumes.}

Fig \ref{Volume} shows the scaling of the percentage volume change 
(going from ideal to relaxed structures) with supercell size, whilst 
the right panel of Fig \ref{Shells} shows the volume changes scaled 
with the number of shells relaxed in the 216 atom supercell.  For the 
antisites and \IiIs a linear fit is again 
rather good, but for all of the other data sets a higher order 
term is clearly present.  We have again tried adding terms in 
$L^{-n}$, $n=2$, 3 and 4.  For both supercell and shell scaling a 
$L^{-2}$ is best.  For supercell scaling $\overline{Q_3}$ is 42 times 
larger than $\overline{Q_2}$ and $\overline{Q_4}$ is 124 times larger.  
The curves plotted in Fig \ref{Volume} are thus quadratics, with the 
$y$-axis intercepts giving the volume change expected for a lone 
defect in an infinite supercell.  These scaled volumes are shown in 
table \ref{Final Ed}.  For shell scaling (Fig \ref{Shells}) 
$\overline{Q_3}$ is 1.4 times larger than $\overline{Q_2}$ and 
$\overline{Q_4}$ is 2.7 times larger.  The quadratic fittings are 
shown as dashed lines in the right hand panels of \ref{Shells}.  The 
$y$-axis intercepts this time give the volume change expected if an 
infinite number of shells were relaxed.  They are shown in table 
\ref{Vol table}, where the equivalent values for supercell size 
scaling are added for comparison.  The two sets of values agree very 
well, indicating that at least the breathing modes in the infinite 
limit are unaffected by charge multipole interactions.

\section{ \label{scale problems}Why some states scale better than 
others.}

\subsection{\label{JT effects}Jahn-Teller active defects.}

\begin{figure}
\epsfxsize 8.5truecm
\epsfbox[15 26 817 575]{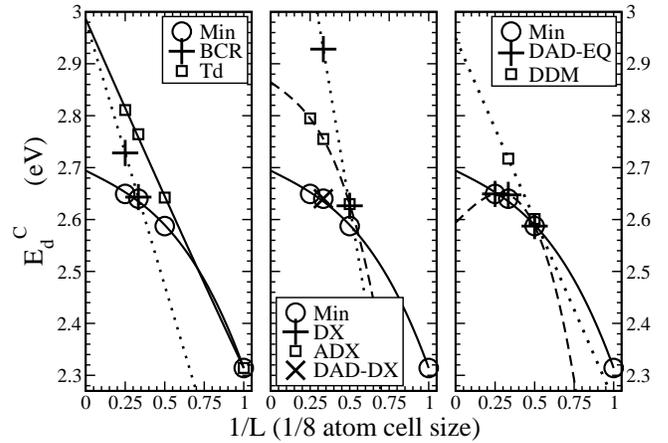}
\caption{\label{JTfig} 
Scaling the formation energies for various Jahn-Teller structures of 
\IP.  In all panels the global minimum is shown (Min).  In addition we show the formation energy when only breathing mode relaxations are allowed (Td) and non-symmetry restricted relaxed formation energies for the BCR, DX, ADX and DDM structures, plus the DAD structure with (DAD-DX) and without (DAD-EQ) an additional DX like distortion.  (See main text for descriptions.) Fits to equation \ref{scale_eqn} are solid when the structure is stable in four 
cells or dashed when it is only stable in 
three.  Dotted lines are linear fits when the structure is 
stable in only two cells.}
\end{figure}

It is very clear from table \ref{Final Ed} that some formation energies 
scale better than others.  When considering the scaling for the neutral 
vacancies \cite{NeutralV} we noted that the scaling becomes more 
difficult to do reliably for strongly Jahn-Teller active defects such 
as \VP.  For a Jahn-Teller active defect there is a 
partially filled degenerate state at the Fermi level, which the 
Jahn-Teller theorem \cite{Jahn-Teller} says will be lifted by symmetry 
reducing relaxations (if no other effect achieves this first).  This 
leads to poor scaling since the symmetry of the most stable relaxed 
structure can vary with supercell size, so that data points from some 
cells scale differently to those from others.  In order to get good 
error bars for scaled formation energies each possible reduced 
symmetry structure must be scaled separately.  Amongst the current 
defects there is a further example of a Jahn-Teller active defect, 
\IP.  The distortions here are much weaker, so the error bar is 
$\pm0.06$ eV even when symmetry differences are ignored, which is 
still reasonable.  Never-the-less, we have done a search for the 
various stable and metastable structures in the four supercells.  
Their various formation energies are shown scaled in Fig \ref{JTfig}, 
together with a) the scaling of the lowest lying formation energy 
irrespective of symmetry (labelled Min) and b) the 
formation energy when only $\hbox{T}_{d}$ symmetry breathing mode 
relaxations are permitted (labelled Td).

In the 8 atom cell the lowest lying structure has 
the full $\hbox{T}_{d}$ symmetry of the unrelaxed antisite.  Indeed, this is the only stable 
structure we find in this cell.  In the larger cells relaxation 
breaks the $\hbox{T}_{d}$ symmetry to give (primarily) $\hbox{C}_{3v}$, 
$\hbox{D}_{2h}$ or $\hbox{C}_{2v}$ point groups at the defect site.  
$\hbox{C}_{3v}$ symmetry is reached by the antisite atom moving either
towards the midpoint of three of its nearest neighbours (away from 
the fourth) in a DX like structure (DX in the figure).  
or towards one neighbour and away from the other three - an anti-DX (ADX) 
structure.  $\hbox{C}_{2v}$ arises when the antisite moves 
towards one pair of neighbours and away from the other pair,  
i.e. towards a bond centre site (BCR).  The degeneracy can also be lifted 
when the antisite stays still with the four antisite-neighbour 
distances equal, but the angles between the bonds changes.  
$\hbox{D}_{2h}$ structures occur if the neighbours rotate to form two 
opposing pairs of either shorter or longer neighbour-neighbour 
distances (DDM - double dimer or DAD - double antidimer, 
respectively).
In the 64 and 512 atom cells the lowest lying structure is a DAD structure with two neighbour-neighbour distances respectively 5\% and 8\% shorter 
than the others.  The most stable structure in the 216 cell is a 
7\% DAD-like structures but with an additional 4\% DX-like distortion, although a 4\% BCR 
structure and a 7\% pure DAD structure (with no DX component) both come a close second. (The distortion quoted for the BCR, DX and ADX structures is the \% variation in antisite-neighbour distances.) In the 512 atom cell the potential energy surface for small (up to $\sim$2\%) DX distortions from the DAD structure is also very flat.  Overall, these results suggest that a lone \IPs in an infinite supercell would have a DAD structure with a formation energy lying about 0.4 eV below the 
$\hbox{T}_{d}$ structure found when only breathing mode relaxations are allowed, and 0.1 eV below the  formation energy found by scaling the minimum formation energy irrespective of Jahn-Teller structure.

The changes in relative stability of the different structures are  
due to one or a combination of two things: a) 
stabilizing/destabilizing dipolar, quadrupolar or higher interactions, which
can in certain cases lift the symmetry without distortion, (in the 8 
atom cell for example,) or favour certain Jahn-Teller structures over 
others.  These effects become weaker as the cells grow.  b) The lack 
of shells of atoms in the smaller cells to absorb the elastic strain, 
which favours more symmetric structures.
In the right hand panel of Fig \ref{Shells} the variation in the 
degree of Jahn-Teller distortion for the ADX and DAD structures was plotted 
versus the number of shells permitted to relax within the 216 atom cell. 
For the ADX structure there is virtually no variation at all and the 
same was previously\cite{NeutralV} found for \VP.  For the DAD structure, on the other hand, a rather strong variation is found. This suggests that elastic 
effects are involved for some local distortion symmetries but not for others, thus further complicating the possible variations in lowest symmetry structure with supercell size.
At least for the ADX structure the charge multipolar 
interactions act essentially in competition with the normal 
Jahn-Teller mechanism, making it uncertain if the correct structure 
has been found at all for smaller supercells.  It is sometimes pointed 
out (Ref [14
] and elsewhere) that one way around 
this problem would be to use k-point integration at the $\Gamma$ point 
only, since this restores the degeneracy of the degenerate levels 
(prior to relaxation).  However, since for InP (and doubtless many 
other materials) the $\Gamma$ point does not give sufficiently 
converged formation energies even in the 512 atom supercell this will 
simply result in unconverged results: errors arising from the use of 
just the $\Gamma$ point can be tenths of eV, often the same size or 
larger than the splittings between various stable and metastable 
Jahn-Teller distorted structures.  Here we study the relative 
stability of all the possible Jahn-Teller distorted structures with 
converged k-point grids and then scale them to predict which structure 
will be most stable in the infinite supercell limit, where these 
spurious degeneracy lifting interactions become zero.  

\begin{figure*}
\epsfxsize 13.0truecm
\epsfbox[3 158 836 587]{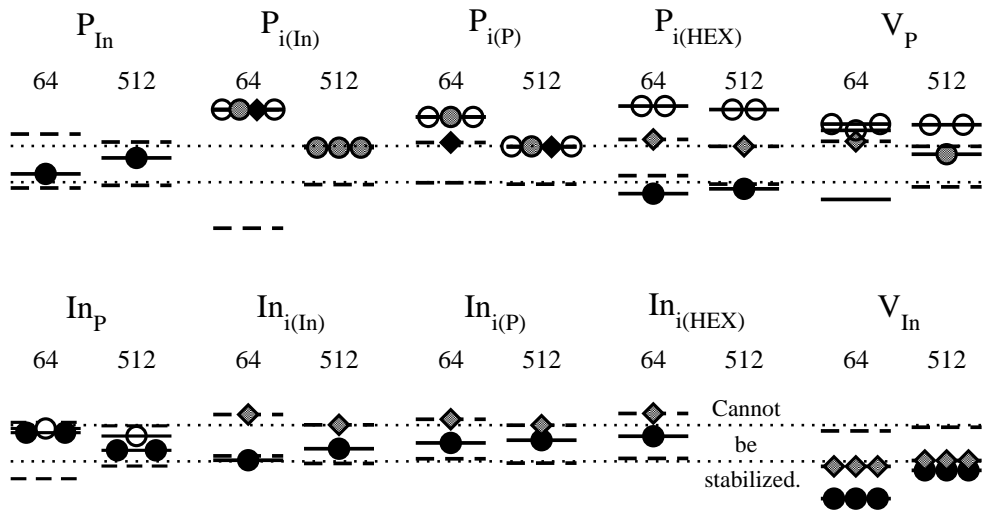} \caption{Defect levels in or near the band gap, 
shown at the $\Gamma$ point for neutral native defects in the 64 and 
512 atom supercells.  Dotted horizontal lines mark the bulk LDA valence 
and conduction band edges.  (0.6771 eV apart.) Dashed lines indicate 
the shifted bulk and valence band edges in the defect cells, solid 
lines indicate localized defect levels.  Black, grey and white symbols 
are (respectively) filled, half filled and empty states.  Circles 
indicate localized states, diamonds delocalized (band) states.  
Filled valence band and empty conduction band states are not 
marked with symbols.}
\label{Bands}
\end{figure*}

\subsection{Defect level dispersion and defect states outside the bandgap.}
\label{nongap state scaling}

The scaling is also rather bad for the 
interstitials, even though no Jahn-Teller behaviour is anticipated or detected 
what-so-ever. One might have expected that the reason for this was the defect level dispersion which was also identified in section \ref{Scaling form} as a possible source of the slight uncertainty in the correct form for the scaling equation. The dispersion can lead to errors  in the formation energies of individual defects and since it should decrease exponentially with cell size it could affect the error bars on the scaled infinite supercell formation energies. In a fully occupied defect level the dispersion should have no direct effect as long as the defect level lies within the gap at all k-points, since the average energy of the band should equal the energy of the level in the absence of dispersion. On the other hand, the dispersion can lead to hybridization of the defect level with conduction band states of the same symmetry, thus artificially lowering the mean value of the defect level and hence of the defect formation energies in the smaller cells. (The same holds for empty defect levels hybridizing with valence band states, allowing the latter to be lowered in energy.) For partially filled defect levels the effect upon the formation energy is more direct, since only the lower parts of the defect level (band) will be filled, again leading to too low a value for the formation energy. However, the connection between the amount of defect level dispersion and the size of the errors in the formation energy in a particular supercell is not simple. Indeed, checks on the bandwidth of the defect levels find no correlation at all between them and the scaling error bars: for example the unrelaxed structures for both \PIs and \IiIs have a defect level in the lower part of the gap with a rather large dispersion ($\sim$0.6 eV in the 64 atom supercell) and yet the scaling error bar for \PIs is 0.02 eV whilst that for \IiIs is 0.35 eV. 

The main difference is actually that \IiIs also has a partially filled defect level {\it inside} the conduction band, resulting in an electron occupying a delocalized state at the conduction band edge. This occurs for all of the defects which have poor scaling error bars. This is seen clearly in Fig \ref{Bands} where we present 
the level diagrams for all the neutral native defects, shown at the 
$\Gamma$ point in the 64 and 512 atom cells.
While the antisites have no delocalized states, the interstitials 
always have at least one electron occupying a state at the conduction 
band edge.  For phosphorus at the tetragonal sites not one but three 
electrons lie above the conduction band edge, which may be why  
they are not stable at all in the larger supercells.  
The level diagrams for the vacancies seem at first glance to contradict 
the rule: \VPs has one delocalized electron in the smaller cells and 
\VIs has three delocalized holes in all four cells, despite the fact 
that the scaling is rather good for these defects.  This is itself a 
finite size effect, however.
For \VPs a triply degenerate defect level lies just above the 
conduction band edge.  This level Jahn-Teller splits, with one state 
moving downwards.  In the smaller cells it does not make it to the 
band gap so the metastable distorted structures found must arise from 
hybridization between the localized Jahn-Teller split levels and levels at 
the conduction band edge.  For the largest cells, however, the lower 
Jahn-Teller split level drops into the gap, so that the charge state 
becomes stable.  In the case of \VIs we again see a triply degenerate 
localized defect level, this time inside the valence band leaving  
three holes at the valence band edge, and even in the 512 atom cell 
this is all we see.  However, as the supercell size grows the defect 
level moves rapidly towards the valence band edge.  Clearly, for an 
isolated defect in a real material we would expect this 
level to lie inside the band gap, even though the 512 atom cell is not 
sufficiently large to show it. This again underlines 
the need for using comparison and scaling rather than just large 
supercells.  (Similar effects are visible for \PiIs and \PiP, although they remain 
unstable.)

\section{Possible charge states across the band gap.}
\label{Non-neutral}

Although we only consider the neutral defects in 
this paper, the level diagrams in Fig \ref{Bands} also give a rough 
indication of which charge states are most likely in different parts 
of the band gap.  For the indium interstitials the +1 charge state 
seems the most stable in the upper part of the band gap, but two 
transition levels will lie in the mid-gap, giving a +3 charge state 
for p-type material.  For \PiHs we expect a +1 charge state across 
most of the gap.  With increasing supercell size a second defect 
level approaches the band edge from below so there may be one or two 
transition levels near the bottom of the gap.  This is a results which 
is only apparent when one compares or scales the results from 
different supercell sizes: even in the 216 or 512 atom cells it is 
not apparent.  \PIs has a single, filled level in the 
upper half of the gap, suggesting 0 for n-type material and +1 or +2 
otherwise.  \IPs has a three-fold degenerate level containing 4 
electrons in the middle of the gap.  A large number of transition levels across the entire gap 
are thus expected, ranging from -1 or -2 in n-type material 
to potentially even +4 for strongly p-type material.  \VPs should 
have two transition levels in the upper half of the band gap, +1 and 
-1 being the most stable charge states in strongly p and n-type 
material respectively.  For \VIs we expect the -3 charge state to be 
stable from the mid-gap upwards.  However, the movement of the 
threefold degenerate defect level up into the band gap (section 
\ref{nongap state scaling}) leads us to expect six transition levels 
all lying near (above or below) the valence band edge.  The 
most stable charge state at the valence band edge itself could be 
anything from 0 to +3.

\begin{figure}
\epsfxsize 6.5truecm
\epsfbox[0 0 340 231]{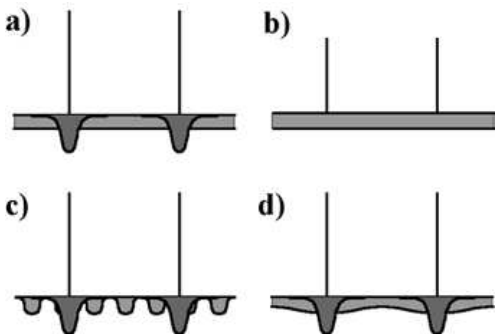}
\caption{\label{Qjellium}
Schematic diagrams for the electrostatic origin of linear 
formation energy error scaling.  a) A positively charged defect, 
with a positive core, localized electrons in a defect level and 
compensating jellium.  b) Subsystem leading to linear errors: weaker 
effective core charge plus jellium.  c) The effective situation for 
\IiP: core charge, two localized defect level electrons and one 
delocalized electron at the conduction band edge.  d) The effective 
situation for the unrelaxed \IP: core charge, four localized defect 
level electrons and two (partially) delocalized defect level 
electrons.}
\end{figure}

\section{The origin of the linear scaling term for neutral defects.}
\label{Linear}

We expect linear scaling terms to arise from two sources: from 
incomplete relaxation (elastic errors) and from electrostatic errors.  
For electrostatic errors linear scaling comes \cite{MP} from the 
monopole term in the multipole expansion of the electrostatic 
interactions: roughly speaking, it is the Madelung energy of the 
localized charge on the defect interacting with the compensating 
jellium background in the image cells.   This is shown schematically 
in Fig \ref{Qjellium}.  Part a) shows the typical 
situation for a positively charged defect: An infinite array of very 
tightly localized positive core charges (one in each periodic image of 
the supercell) are only partially compensated by the localized 
electrons in the (less tightly) localized defect levels.  The remaining 
charge is compensated by jellium.  Makov and Payne\cite{MP} extracted 
the part of the system shown in Fig \ref{Qjellium} b) consisting of an 
infinite array of positive delta functions interacting with  
jellium and showed that is gives rise to a formation energy error 
which is linear in the supercell size, with a strength proportional to 
the square of the charge on the defect.  This error should therefore 
be absent for unrelaxed neutral defects.  The fact that we see a clear 
linear contribution to the unrelaxed formation energy of, for example, 
the neutral \IPs and tetrahedral In interstitials is thus somewhat 
unexpected.

\begin{figure}
\epsfxsize 8.5truecm
\epsfbox[19 26 823 575]{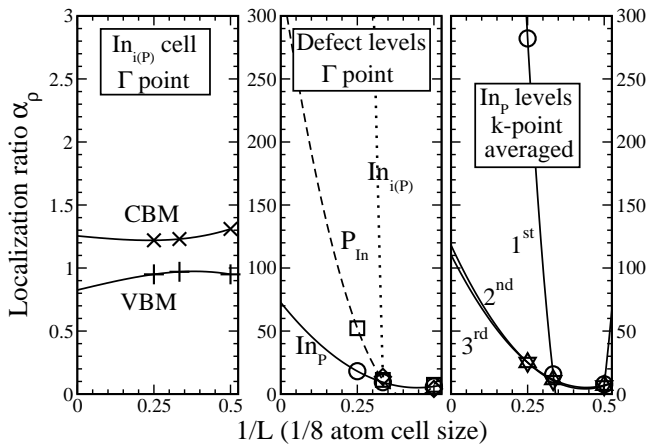}
\caption{\label{Loc} 
Localization of the Kohn-Sham levels corresponding to (left panel) the 
valence and conduction band edges at the 
$\Gamma$ point in the cell containing \IiP; (middle panel) the defect 
levels in the band gap for \IiP, \PIs and \IP, calculated at the 
$\Gamma$ point; (right panel) the three \IPs defect levels averaged 
over the whole Brillouin zone.  See text for details.  Lines are 
simple quadratic fits for guidance: the correct form for this scaling 
has not been investigated.}
\end{figure}

The explanation lies in the localization/delocalization of the states 
which become occupied and unoccupied when the cell contains a neutral 
defect.  In the cases of the neutral interstitials and vacancies one 
or more defect level(s) lie outside the bandgap for all or some 
supercell sizes.  As a result there are either electrons in delocalized 
band states at the LDA conduction band edge or holes at the valence band 
edge.  This is shown for the example of \IiPs in Fig \ref{Loc}.  
To quantify the localization of a particular Kohn-Sham eigenstate we 
start by projecting it onto Wigner-Seitz cells\cite{Wigner-Seitz} 
around each atom.  We then select some radius $r_{\hbox{l}}$ around 
the defect and find the average projection per atom $\rho_{>}$ over 
the part of the cell where $r > r_{\hbox{l}}$, and similarly the 
average $\rho_{<}$ for the $r < r_{\hbox{l}}$ region.  The 
localization in then given by the ratio
\begin{equation}\label{loc ratio}
\alpha_{\rho}\!\left(r_{\hbox{l}}\right) = \frac{\rho_{>}}{\rho_{<}}
\end{equation}
\noindent For an isolated defect in an otherwise perfect infinite 
lattice, $\alpha_{\rho}\rightarrow\infty$ as 
$r_{\hbox{l}}\rightarrow\infty$ if the state is localized, since 
$\rho_{>}\rightarrow0$.  On the other hand, 
$\alpha_{\rho}\rightarrow1$ if the state is delocalized as the two 
averages $\rho_{>}$ and $\rho_{<}$ then tend to the same value.

Here we have finite $L$ sized supercells, so  we choose 
$r_{\hbox{l}}$ such that $\rho_{>}$ is calculated over the  
outermost complete shell of atoms around the defect, plus the atoms 
in the cell corners.  $\rho_{<}$ is 
then calculated over the remaining 2, 4 and 6 complete shells in the 64, 
216 and 512 atom cells respectively.  (The 8 atom cell is too small 
for this analysis.) Supercell scaling of 
$\alpha_{\rho}\!\left(r_{\hbox{l}}\!\left(L\right)\right)$ then has the 
same behaviour as for the isolated defect.  Thus, in Fig \ref{Loc}, 
$\alpha_{\rho}$ is plotted against one over the cell size for various 
states of interest.  The left panel shows the localization scaling of 
the states at the LDA valence band maximum (VBM) and conduction band 
minimum (CBM) for cells containing an unrelaxed \IiP, whilst the 
middle panel (with a very different vertical scale) shows it for the 
mid-gap defect levels of both \IiPs and the antisites.  The well 
behaved \PIs is included for comparison. These localizations, like the 
level diagrams in Fig \ref{Bands}, have been calculated at the 
$\Gamma$ point, but using fully k-point converged charge densities.  For \IiPs 
the band states are delocalized and the defect level is localized.  
The VBM and defect levels are filled, whilst the delocalized CBM level 
is half filled.  Hence adding the defect to the cell has added 2 
localized electrons and one delocalized electron in the vicinity of the 
band gap.  Hence, we have an electrostatic situation like that 
shown schematically in Fig \ref{Qjellium} c).  To a first 
approximation we can replace the charge density of the delocalized 
electron by its average value, thus recovering the situation of 
Fig \ref{Qjellium} a).  We thus predict a linear term in the formation 
energy scaling.  However, the distribution of the delocalized charge is 
in reality far from uniform on the scale of the atomic spacing, so it
is not clear what the prefactor and effective charge should be.  
Hence, without performing the detailed mathematical derivation 
required (which lies beyond the scope of the current paper) we cannot 
predict what gradient the linear term should have.

The case of \IPs is different.  There are no electrons in 
delocalized conduction band states or holes in the valence band.  
However, the localization of the defect level at the $\Gamma$ point is 
rather weak, certainly compared to \PIs and \IiP. The defect level of \IPs is 
three-fold degenerate and partially filled.  Away from the $\Gamma$ 
point this degeneracy is split by the interaction of the defect with 
its images, leading to different dispersions of the defect levels in 
different directions in k-space, with one level having lower energy 
(away from $\Gamma$).  At most k-points this state is thus completely 
filled, whilst the other two are partially empty.  Hence, averaging 
the Wigner-Seitz projection over the whole Brillouin zone (using 
8x8x8, 4x4x4 or 2x2x2 point Monkhorst-Pack grids in the 64, 216 and 
512 atom cells respectively) the completely filled level becomes fully 
localized.  (Right panel of the figure.) The other two levels are more 
occupied at the origin in k-space than elsewhere and hence in real 
space are only partially localized.  Clearly the Jahn-Teller structural 
relaxations are required in order to properly localize these states.  In the unrelaxed 
structure we effectively have two localized electrons and two 
partially delocalized ones in the middle of the band gap, leading to the 
spatial charge distribution shown 
schematically in Fig \ref{Qjellium}d).  Electrostatically, this again 
behaves to first order like a jellium background charge surrounding a 
positively charged, localized defect, resulting in the linear scaling 
observed.  As with the interstitials, predicting the specific gradient 
expected must be left to future work.  Here, we simply note that this 
mechanism should also affect the unrelaxed formation energies of other 
defects with partially filled degenerate levels in the band gap.  It 
seems reasonable to presume that it is involved with the neutral \VIs 
and perhaps \VP, although they also have linear terms arising from 
partially filled band states.

\begin{figure}
\includegraphics[height=6.0truecm]{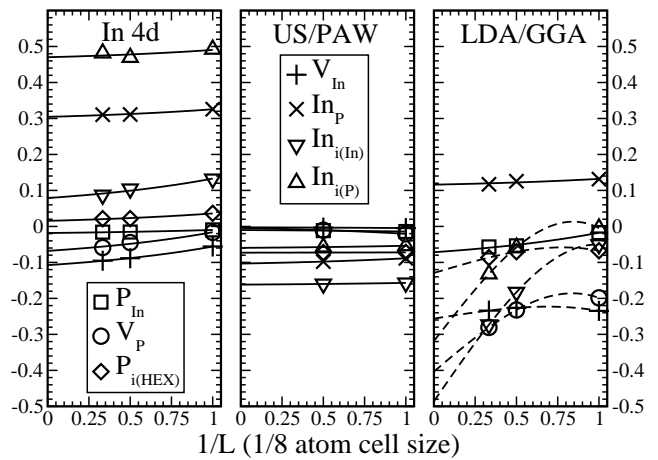}
\caption{\label{Errors} 
Scaling of errors from other sources. Left panel: the change in formation 
energy when the In 4d electrons are treated as core rather than 
valence - i.e. the formation energy using an US-PP with In 4d 
treated as valence minus the formation energy found earlier with the 
In 4d as core. Centre panel: the change in formation energy when PAW is used 
rather than US-PP (both with In 4d 
are treated as valence.) Right panel: the change in formation energy 
when GGA is used instead of LDA. Solid line fits are to Eq 
\ref{scale_exp}, dashed lines are to Eq \ref{scale_eqn}.}
\end{figure}

\section{Other Sources of Error.}
\label{Other errors}

Finite size errors are not the only types of errors 
present in these calculations: errors also arise from the truncation of the basis set and the k-point integration as well as from the 
use of both pseudopotentials and LDA. Furthermore, memory size limitations 
forced us to use pseudopotentials in which the indium 4d shell is treated as core 
rather than valence. Although the central aim of this paper is to 
study the treatment of finite size errors it is still informative to 
estimate the size of these other sorts of errors for comparison. 
In principle, errors arising from the pseudopotentials and the LDA should be 
independent of supercell size, though some short range
dependence may still be anticipated since changes may affect the amount 
of defect band dispersion. This contribution should disappear 
exponentially with supercell size. 

In fig \ref{Errors} we show the change in the unrelaxed formation 
energies when (left panel) the In 4d electrons are treated as core rather 
than valence, (centre panel) the US-PP are replaced by the Projector Augmented Wave method\cite{PAW} (PAW) 
or (right panel) the LDA exchange correlation 
functional is replaced by the Generalized 
Gradient Approximation (GGA) of Perdue and Wang\cite{GGA}. These formation energy differences are 
shown as a function of supercell size for the seven stable point 
defects. (\VI, \IP, \IiI, \IiP, \VP, \PIs and \PiH.) For the GGA and for LDA with the In 4d as valence a planewave 
cutoff of 200 eV gives the same level of accuracy as it does with the In 4d as core. For PAW on the other hand a 
planewave cutoff of 300 eV is required for the same accuracy level.
The relaxed lattice constants and bandgaps also change. For LDA 
with In 4d as valence we find 5.833 \AA~and 0.581 eV respectively, for LDA with PAW we find 5.830 \AA~and 0.597 eV whilst for GGA 
(with US-PP and In 4d as core) we find 5.956 \AA~and 0.473 eV.

The errors coming from the pseudopotentials (left and centre panels of 
Fig \ref{Errors}) are, as anticipated largely independent of the 
supercell size. The slight dependence is well described by a two parameter fit to the general 
exponential
\begin{equation}\label{scale_exp}
E_{\rmd}^{\rmC}\!(L) = E_{\rmd}^{\infty} + a_{e}\left(\exp\left(L^{-1}\right)-1\right)
\end{equation}
\noindent (Solid lines in the figure.) Changing the In 4d electrons from core to 
valence has, not surprisingly, only a fairly small effect (well below 0.1 eV) 
upon the phosphorus related defects, 
but a much more significant effect upon the indium related ones, 
particularly \IiIs and \IPs where the effect is on the same 
order of magnitude as the finite size errors. PAW produces more accurate results than any
pseudopotential method since it reconstructs the exact valence wavefunction with all nodes in the core region. The replacement of
US-PP with PAW produces only small ($\Or$(0.1 eV) or less) changes in the formation energies with virtually no 
size dependence: the largest difference between the correction in the 
8 atom cell and that in the 64 atom cell is only 0.01 eV, so the US-PP to 
PAW changes have not been calculated in the 216 atom cell. This demonstrates that 
the widely used US-PP\cite{USPP} are perfectly reasonable for this type of calculation. It should also be noted that in most cases the small correction when PAW is introduced partially cancel that made when moving the In 4d electrons from core to valence, at least for the examples studied here.

The errors arising from the exchange-correlation functional have two main forms. Firstly, the band gap is (usually) strongly reduced compared to experiment. (For InP GGA does worse than LDA once lattice parameter optimization has been included.) This reduction leads to ambiguities in the definition of the formation energy for charged defects, in turn leading to large uncertainties in predictions. For some semiconductors the bandgap can even be reduced to zero making the material appear metallic and dramatically altering the properties of many defects. However, neither of these effects occurs here, since we consider neutral defects and obtain a non-zero band gap. Never-the-less, a second exchange-correlation related error is present: LDA overbinds all bonds, moving some defect formation energies up and others down. Hence to assess the errors involved we here compare with GGA, which is known to have the opposite effect: underbinding. An exact solution to the DFT formation energies would lie somewhere in between - and probably closer to the LDA results since LDA gives a better lattice constant for bulk InP. We note also that we have not allowed spin polarization in our calculations, using LDA instead of the Local Spin Density Approximation (LSDA). However, for most of our defects the use of LSDA would simply cause further finite size errors since it would introduce spurious magnetic interactions between the defect and it's PBC images. The exceptions occur for the Jahn-Teller active defects ({\IP, \VPs and \VI) where the degeneracies can in certain cases be lifted by Hund's rule couplings. However, this is only important for materials\cite{SiC} with more tightly localized bonds or dangling bonds than we have here, so we may safely omit it.

The results from replacing LDA with GGA are shown in the right panel of Fig \ref{Errors} and are somewhat surprising: there 
is a clear finite size error associated with the choice of 
exchange-correlation functional for certain defects, namely \IiP, \IiI, \VPs 
and \PiH. The antisites on the other hand 
show no significant supercell size dependence. (\VIs shows only a small 
size dependence but is still fitted better by a polynomial than by an exponential.) The defects for which the LDA/GGA error difference depends on supercell size are those 
for which the defect levels lie outside the 
band gap. This indicates that the spurious 
delocalized states are treated differently by different exchange correlation functionals. 
This is confirmed by checking the degree of localization of the 
electrons/holes at the band edges: when GGA is used the values of $\alpha_{\rho}$
are 20 to 400\% larger than with LDA. These changes are 
linked to the change in band gap and also to the fact that the defect bands move
closer to the band edges (at least near the $\Gamma$ point) for these particular defects.
Hence we may expect large, supercell size dependent errors of $\Or$(0.5 eV) from the 
choice of exchange correlation functional when partially filled 
defect bands lie outside the bandgap but we expect smaller $\Or$(0.1 eV) size independent errors when the defect levels lie within 
the band gap. 

Table \ref{Error comparison} compares the size of the errors arising from different 
sources. The basis set errors are estimated from the difference in formation energy in the 64 
atom supercell when the planewave cutoff is raised from 200 eV to 400 eV. The k-point errors are different in each 
supercell: the errors shown in the table are the largest occurring for any of the supercells used, estimated as the 
difference between the mean value ${\overline{E_{\rmd}^{\rmC}}}$ and the value  $E_{d}^{C}\!\left(N_{\hbox{max}}\right)$ obtained 
using the largest k-point grid actually calculated for the defect and supercell in question.
All other sources of error - such as defect band dispersion etc - are contained within the finite size errors 
shown.
Most of the errors listed are on the 0.1 eV scale or below, which in practical calculations 
is usually acceptable. The finite size errors and some of those arising from 
treating the In 4d electrons as core are larger, lying on the 
$\sim$0.5 eV level. We note, however, that for charged defects we 
anticipate even larger finite size errors - up to 1-2 eV in many cases\cite{Us future} - whilst we do not anticipate the errors from the 
In pseudopotentials being significantly different from those here. 
The presence of defect level dispersion effects is confirmed by the existence of
the exponentially shrinking supercell size dependence in the errors related to pseudopotentials and to the LDA for the antisites. These results also confirm that they are short ranged. Their energy scale is rather hard to estimate directly, but the size of the exponential components in Fig \ref{Errors} suggests that the errors involved are only on the 0.01-0.1 eV scale even for the 8 atom cell. An alternative estimation comes from the error bars on the scalings themselves (see table \ref{Final Ed}) which are 0.01-0.35 eV, only a small part of which comes from the defect level dispersion.

The fact that the errors coming from the pseudopotentials have only a small and exponentially decaying cell size dependence means that it is perfectly reasonable to make the approximations we needed to make in order to be able to do calculations in sufficiently large supercells to correctly assess the finite size errors. The amplitude
and sign of these non-size dependent errors can be calculated separately in a smaller cell - the 64 atom cell for example - and simply added or subtracted from the final finite size scaled results in order to produce much more accurate and reliable 
defect formation energies than those normally published. This has been done for the 
LDA formation energies of the present examples in the last column of table \ref{Error comparison}, but is equally valid for the errors in the relaxed formation energies and those in the GGA (at least if no delocalized hole/electron states have appeared at the valence/conduction band edges.) 

\begin{table}
\caption{\label{Error comparison}ÊComparing the size of the errors 
arising from the various different approximations. i) Finite size errors from the supercell 
approximation (shown for the 64 and 512 atom supercells.) ii) The 
treatment of the In 4d electrons as core (scaled to an infinite cell.)  iii) US: the use of ultrasoft 
pseudopotentials, compared to PAW (scaled to an infinite cell.) iv) The LDA, compared to GGA, (scaled to an infinite cell. For the cases in which a finite size term appears in the LDA vs GGA error a more valuable comparison is of the errors in individual cells, so the values in the 64 and 512 atom cells are then given in brackets.) 
v) Basis set truncation (in the 64 atom cell) and vi) k-point integral truncation. (Shown for the cell in which it is worst for the defect in question.) vii) ``Final'' shows the scaled LDA formation energy when the 
pseudopotential errors are accounted for.
(All energies in eV.)}
\begin{center}
\begin{tabular}{llllllll}
\hline\hline
Defect & Supercell   & In 4d  &US         &LDA             &Basis    &K-grid  &Final   \\\hline
\VI    & 0.40/0.20   &0.11    &0.00       &0.25(0.24/0.23)  &0.003    &0.003     &4.84    \\
\IP    & 0.41/0.21   &0.30    &0.10       &0.12             &0.003    &0.0007    &3.57    \\
\IiI   & 0.40/0.25   &0.08    &0.16       &0.48(0.18/0.32)  &0.01     &0.007     &4.67    \\
\IiP   & 0.55/0.33   &0.47    &0.06       &0.32(0.05/0.17)  &0.02     &0.007     &4.56    \\
\PI    & 0.03/0.03   &0.02    &0.01       &0.07             &0.01     &0.002     &2.46    \\
\VP    & 0.12/0.03   &0.07    &0.00       &0.40(0.23/0.31)  &0.004    &0.001     &2.93    \\
\PiH   & 0.06/0.02   &0.02    &0.07       &0.13(0.07/0.10)  &0.04     &0.009     &4.85    \\
\end{tabular} 
\nl
\end{center}
\end{table}

\section{Conclusions.}
\label{Conclusions}

We have studied the finite size errors which occur when the supercell approximation is used in the calculation of the
formation energies and structures of point defects in semiconductors, using the neutral native defects of InP as an 
example. We have calculated the relaxed and unrelaxed formation energies using planewave ab 
initio DFT in simple cubic supercells containing 8, 64, 216 and 512 atoms - the largest currently computable.  To examine 
and correct for these errors we have  used  finite size scaling with inverse supercell 
size, which we consider to be the most reliable and accurate way to treat the finite size supercell approximation 
errors. Unlike other methods this does not rely upon any modelling or assumptions about the errors, other than 
that they are primarily long ranged (polynomial rather than exponential) and decrease with increasing supercell size. This method requires the results of calculations in at least 3 supercells, and at least 4 if we are to have an idea of the accuracy of the resulting scaled energies. Hence for some difficult cases in which the 8 atom cell is simply {\it too} unreliable it may occasionally be necessary to use supercells with up to 1000 atoms.

Three sources of finite size error have been examined: in the 
case of relaxed formation energies there are elastic errors due to the 
finite volume available for relaxation.  We showed that, as they 
should, these scale linearly with inverse supercell size ($L^{-1}$) 
with very little hint of any higher order error term arising, even when 
Jahn-Teller distortions are taken into account.  The second type of error is the dispersion of the defect levels, which has only a relatively small effect upon the formation energies, at least when the defect levels are completely filled. These effects appear to shrink exponentially with increasing supercell size, as anticipated, but appear to slightly increase the uncertainties in the final scaled formation energies. The third source of 
error is much more significant and arises for both relaxed and unrelaxed formation energies. It  is 
due to charged multipole interactions between a defect and its images 
in the PBCs.  We have shown that these errors are present and not 
always negligible, even for neutral defects.  Linear errors can arise 
if fully or partially filled defect levels lie outside the band gap in the neutral charge state, leading to delocalized holes at the valence band 
edge or electrons at the conduction band edge. Linear errors can 
also appear in unrelaxed formation energies due to the way in which the defect/image 
interactions lift degeneracies for partially filled degenerate defect 
states within the band gap.  Both of these ways for linear scaling 
errors to arise can apply for other non-neutral defects.  
They indicate that the calculation 
is in some way unphysical, for example that the charge state involved is not actually stable. However, in practise it is often necessary to calculate formation energies of such unstable charge states in order to check which transitions levels do actually lie inside the gap. It is thus important to be aware that large finite size errors, such as those reported here, can occur even in calculations for neutral defects, as this can lead to transition levels calculated in individual supercells (without scaling) to appear to lie inside the gap when they should lie outside and vice versa.
Indeed, electrostatic errors are still present even for physically reasonable 
cases, such as \VPs and (probably) \VI, due to defect states which enter the gap as the cell size grows.
They are even present and may remain significant for defects such as the neutral \PIs which has all of its defect levels within the gap in all supercells.
This is because neutral defects in crystalline solids still have 
higher order charge multipole moments, especially if the system is 
made up of more than one type of atom with different 
electronegativities.  We have shown that for unrelaxed formation 
energies these errors scale as the inverse cube of the supercell size 
($L^{-3}$).  For relaxed formation energies this is almost certainly 
the case also, although we do find possible indications that the leading non-linear error term 
may sometimes scale as the inverse square ($L^{-2}$).  Further work with more 
defects is needed to confirm or definitively rule out this possibility.  
It could clearly be answered by removing the 8 atom cell from the scaling and 
replacing it with the 1000 atom cell, but that must wait for improved 
computing facilities.  In the meantime, an answer may be obtained from 
scaling studies of formation energies for charged defects \cite{Us 
future} since both the electrostatic errors themselves and the cross 
terms between them and the elastic errors will then be stronger. 

To summarize: the use of large (up to 500 or occasionally even 1000 atom) supercells with 
finite size scaling has been shown to be a very promising route around 
the errors which arise in the use of the supercell approximation to 
calculate formation energies of defects in III-V (and other) 
semiconductors.  Errors scale with a linear plus a higher order term, most
probably cubic.  We have also found several instances where 
scaling recovers physically relevant results that are even hidden in 
calculations on the 512 atom cell: formation energies which are wrong 
by $\sim\frac{1}{2}$ eV and defect levels which appear inside the 
valence or conduction bands in supercell calculations when they should actually 
lie inside the band gap.

\section*{Acknowledgements}

The authors would like to 
thank A.  H\"oglund for useful discussions, as well as U. Gerstmann 
and his coworkers at the University of Paderborn, Germany.  
The calculations in this paper were performed at Uppsala University and 
at the Parallel Computing Centre (PDC) Stockholm, Sweden.  The authors 
would also like to thank the G\"oran Gustafsson Foundation, the 
Swedish Foundation for Strategic Research (SSF) and the Swedish 
Research Council (VR) for financial support.


\section*{References}

\begin{thebibliography}{9}
\bibitem{DFT} W. Kohn and L. Sham Phys. Rev. {\bf 140} A1133 (1965)
\bibitem{MP} G. Makov and M.C. Payne Phys.  Rev.  B {\bf 51} 4014 (1995)
\bibitem{OtherKorr} P.A. Schultz Phys Rev B {\bf 60} 1551 (1999);
         L.N. Kantorovich Phys. Rev. B {\bf 60} 15476 (1999);
         H. Nozaki and S. Itoh Phys. Rev. E {\bf 62} 1390 (2000)
\bibitem{Matt} M.I.J. Probert and M.C. Payne Phys. Rev. B {\bf 67} 75204 (2003)
\bibitem{NeutralV} C.W.M. Castleton and S. Mirbt {\it Physica B} {\bf 340-342} 407 (2003)
\bibitem{Jahn-Teller} H.A.  Jahn Proc Roy Soc London A Mat {\bf 161} 220 (1937)
\bibitem{Old interstitials} R.W. Jansen Phys Rev B {\bf 41} 7666 (1990)
\bibitem{Old antisites} A.P. Seitsonen, R. Virkkunen, M.J. Puska and R.M. Nieminen Phys Rev B 5253 (1994)
\bibitem{USPP} D. Vanderbilt Phys Rev B {\bf 41} 7892 (1990);
         G. Kresse and J. Hafner J. Phys: Cond. Matt. {\bf 6} 8245 (1994)
\bibitem{VASP} G. Kresse and J. Furthm\"uller Comp. Mat. Sci. {\bf 6} 15 (1996)
\bibitem{ZVP} C.W.M. Castleton and S. Mirbt Phys Rev B {\bf 68} 085203 (2003)
\bibitem{Us future} C.W.M. Castleton and S. Mirbt In preparation.
\bibitem{Monkhorst-Pack} H. Monkhorst and P. Pack Phys. Rev. B {\bf 13} 5188 (1976)
\bibitem{JT and Gamma point} M. Bockstedte, A. Mattausch and O. Pankratov Phys. Rev. B {\bf 68} 205201 (2003)
\bibitem{Wigner-Seitz} A. Eichler, J. Hafner, J. Furtm\"uller, and G. Kresse Surf. Sci. {\bf 346} 300 (1996)
\bibitem{PAW} P.E. Bl\"ochl, Phys Rev B {\bf 50} 17953 (1994); 
         G. Kresse and J. Joubert Phys Rev B {\bf 59} 1758 (1999)
\bibitem{GGA} J.P. Perdew and Y. Wang Phys Rev B {\bf 13} 5188 (1976)
\bibitem{SiC} A. Zywietz, J. Furthm\"uller Phys Rev B {\bf 59} 15166 (1999)
\end{thebibliography}
\end{document}